# CNN Application in Detection of Privileged Documents in Legal Document Review

Rishi Chhatwal, Esq.
Legal
AT&T Services, Inc.
Washington DC, USA
rishi.chhatwal@att.com

Robert Keeling, Esq.
Complex Commercial Litigation
Sidley Austin LLP
Washington DC, USA
rkeeling@sidley.com

Peter Gronvall
Data & Technology
Ankura Consulting Group, LLC
Washington DC, USA
peter.gronvall@ankura.com

Nathaniel Huber-Fliflet
Data & Technology
Ankura
Washington DC, USA
nathaniel.huber-fliflet@ankura.com

Jianping Zhang
Data & Technology
Ankura Consulting Group, LLC
Washington DC, USA
jianping.zhang@ankura.com

Haozhen Zhao
Data & Technology
Ankura
Washington DC, USA
haozhen.zhao@ankura.com

*Abstract*— Protecting privileged communications and data from disclosure is paramount for legal teams. Legal advice, such as attorney-client communications or litigation strategy are typically exempt from disclosure in litigations or regulatory events and are vital to the attorney-client relationship. To protect this information from disclosure, companies and outside counsel often review vast amounts of documents to determine those that contain privileged material. This process is extremely costly and time consuming. As data volumes increase, legal counsel normally employs methods to reduce the number of documents requiring review while balancing the need to ensure the protection of privileged information. Keyword searching is relied upon as a method to target privileged information and reduce document review populations. Keyword searches are effective at casting a wide net but often return overly inclusive results – most of which do not contain privileged information. To overcome the weaknesses of keyword searching, legal teams increasingly are using machine learning techniques to target privileged information. In these studies, classic text classification techniques are applied to build classification models to identify privileged documents. In this paper, the authors propose a different method by applying machine learning / convolutional neural network techniques (CNN) to identify privileged documents. Our proposed method combines keyword searching with CNN. For each keyword term, a CNN model is created using the context of the occurrences of the keyword. In addition, a method was proposed to select reliable privileged (positive) training keyword occurrences from labeled positive training documents. Extensive experiments were conducted, and the results show that the proposed methods can significantly reduce false positives while still capturing most of the true positives.

## I. INTRODUCTION

In United States, companies responding to litigation or a request from an enforcement agency (e.g., Department of Justice or the Securities and Exchange Commission), typically are obligated to produce to the requesting party all non-privileged material relevant and proportional to the legal case [1]. To accomplish this, the company's legal teams most often are tasked with gathering, compiling, and reviewing large volumes of business documents to determine which documents are relevant to the legal case, and then providing copies of those relevant documents in the form of a document 'production.' Corporations can spend millions or tens of millions of dollars to accomplish this very cumbersome, expensive, and legally-required task. This entire process is referred to in the legal industry as electronic discovery (or 'e-discovery') and more specifically, the 'document review' component of e-discovery. Document review requires significant time and resources to meet production schedules established by the legal process. The costs of document review continue to escalate as the volumes of business data continue to grow. In fact, it can be safely stated that for all practical purposes, data volumes are approaching 'infinite' in size, in the sense that it is essentially impracticable to review each document potentially falling within the scope of an investigation or litigation.

When attorneys are reviewing to identify relevant documents, they also typically work to identify which documents contain privileged material. In most cases, any documents containing privileged material can and should be withheld from production. Privileged material in today's legal environment are usually emails and electronic documents, either consisting of communications involving lawyers, or documents prepared at the request of lawyers or otherwise in connection to an actual or anticipated legal matter. Privileged materials are also generally 'protected' from disclosure, by the attorney-client privilege and the work product doctrine. Because of these important disclosure protections, it is critical for lawyers to screen for these types of documents to ensure that they are removed from document productions provided to requesting parties. The accidental production of privileged communications or work product can be potentially devastating to a legal matter; these documents could provide an opposition party with insight into a company's proposed legal strategy, regulatory decision-making process, or internal investigation findings.

With data volumes increasing in legal matters, privilege review places a large burden on legal counsel when performing the document review – further increasing cost. Legal teams employ numerous workflows when undertaking privilege document review. Workflows that employ keyword searching to target potentially privileged documents are commonly used to focus the privilege review population and reduce the volume of documents requiring manual review. Typically, counsel develops a list of potentially privileged keyword terms and applies those terms to the relevant document population to find documents that contain privileged information. These documents are presumed as potentially privileged and are reviewed by counsel to confirm their privilege.

In response to the surging volumes – and incumbent costs – of document review, the legal industry increasingly is turning to machine learning and text classification techniques in search of greater efficiencies and overall accuracy. Most specifically, attorneys and their clients are applying machine learning to text classification models to identify relevant documents in legal cases. Machine learning techniques have been successful at reducing review populations in legal matters when targeting relevant materials [2]. This reduces the time and cost of attorney review. While using machine learning to target relevant content has been embraced by the legal community, there is a stigma that machine learning cannot reliably identify privileged material. Anecdotally, attorneys maintain a belief that a machine learning model is not precise enough to classify privileged material due to the nuance of specific privilege determinations and the relationship-driven nature of communications.

There is little research about the use of machine learning techniques to target privileged information other than Gabriel, et al. [3] and [4]. In these studies, classic text classification techniques are applied to build classification models to identify privileged documents. Because privileged information often occurs in a small piece of text in a document, classic text classification techniques cannot significantly improve the performance of keyword searches [4]. In this paper, the authors report our work in applying Convolutional Neural Networks (CNN) to improve keyword search performance in identifying privileged documents. Our work combines keyword searching with machine learning. For each keyword term, a CNN model is created using the context of the occurrences of the keyword to distinguish privileged occurrences from non-privileged occurrences. Training data for keyword occurrences are generated from labeled privileged and non-privileged documents. Keyword occurrences in privileged documents are not necessarily privileged occurrences. Therefore, we proposed a method to select a subset of most likely privileged occurrences as training data. Extensive experiments have been conducted on three legal document review matters and the results show that machine learning can significantly reduce false positives while still capturing most of the true positives. Popular search terms such as "Legal", "Attorney", "Privilege", and "Counsel" are used in different legal matters to identify privileged documents. The purpose of this work is to build a CNN model for each of these popular search terms and to apply the model to new legal matters in the future.

## II. LEARNING KEYWORD OCCURRENCE MODELS

Keyword searching is a common approach used to target privileged material. A list of keyword terms is created by the legal team and then those terms are searched across the document population to identify documents that contain a term hit. Keyword searching often yields large numbers of 'false positive' documents to ensure the results are comprehensive. Common terms included in a privilege keyword term list are, "privileged" and "legal".

Whether or not an occurrence of a keyword term in a document is privileged depends upon the context of the occurrence. For example, when "legal" occurs in the footer language of an email, the occurrence of "legal" is likely not privileged because, among other reasons, the email footer does not contain the request or provision of legal advice. In addition, many non-attorney employees use default email confidentiality disclosure footers and their documents are typically not privileged. In our work, we train a CNN model for each search term to improve the precision of the search term by using words that occur around the term. In this approach, we create a window for each occurrence of a term using n words before and n words after the occurrence. Namely, each occurrence of a term is converted into a $2 \times n+1$ word text snippet, which is treated as a document. We chose to experiment with the CNN learning algorithm because the documents are short with a fixed number of words and we have a large number of training text snippets.

This problem is similar to the word-sense problem. The same word can have different "senses." For example, "apple" has two senses, a type of fruit and a company name. The words around "apple" in different senses are likely different. Therefore, we should be able to build a machine learning model using the context of those words to distinguish the two senses in our "apple" example. Similarly, most word embedding techniques [5][6] leverage word contexts to convert a word into a numeric vector.

In privileged document review, the cost of manually labeling training data is high and it is typically too expensive to have an attorney evaluate the words around a privilege keyword term to confirm if that occurrence of the keyword term is privileged or not. Therefore, training data for a

keyword occurrence model needs to be derived from whole documents labeled by attorneys. Non-privileged keyword occurrences can be extracted from the keyword occurrences in non-privileged documents. An occurrence of a keyword term in a non-privileged document is always not privileged since the entire document contains no privileged information, therefore training data **for not privileged keyword occurrences is accurate**. We can also extract privileged keyword occurrences from privileged documents. However, an occurrence of a keyword in a privileged document may not be privileged because the document may be privileged for a different reason other than the keyword term – consider the presence of "legal" in the email footer example earlier in this paper. Therefore, **privileged keyword occurrences extracted from privileged documents are less accurate than occurrences from the non-privileged documents but still likely privileged keyword occurrences**. In our experiments, we used these likely privileged keyword occurrences as positive training labels and achieved promising results. In Section 5, we propose a method to select a subset of very likely privileged training documents.

The purpose of determining privileged keyword occurrences is to identify privileged documents. A keyword occurrence model assigns a score between 0 and 1 to each keyword occurrence. The larger the score, the more likely the occurrence is privileged. Each document also receives a score which is the largest score of all occurrences of the keyword present in the document. In our study, a document is considered privileged if at least one privileged keyword occurrence has a high enough score to meet a given recall threshold.

### III. Convolutional Neural Network for Text Classification

In last decade, deep learning has made tremendous progress in both research and applications. Deep learning has been successfully applied to many visual analysis and natural language processing tasks. Convolution Neural Network is a deep learning technique and has been widely used in image processing. Recently, CNN has been adapted to text classification and proven effective in academic research when applied to real world data, such as Yelp and IMDB reviews for predicting customer ratings and also to determine the sentiment of tweets from Twitter [7, 8, 9].

In our work, we used a simple CNN structure introduced in a 2014 study [7]. The simple CNN structure consists of a single one-dimension convolution layer followed by dropout, one-dimension max pooling, and a fully connected layer with binary classification. We chose to use an embedding layer as part of training instead of the pretrained word embedding used in the 2014 study [7]. The parameters for CNN were set as follows.

• Filter Number: 64

• Filter Kernel Size: 2

• Maximum Pooling Size: max (i.e., 1-max)

• Tokenizer Vocabulary Size: 20,000

• Tokenizer Sequence Length: Based on the window size of search term occurrence

These parameters were chosen based on experiments with the legal matter datasets in [10]. In prior research [7, 8, 9, 11], authors have discussed various hyperparameter settings, and our parameter choices followed some of their recommendations, including using 1-max pooling. These hyperparameters settings were fixed across datasets and training set sizes. But the dropout rate was customized for each training set – the dropout rate and epochs parameters were chosen by analyzing the results of a grid search of various combinations and identifying the combination with the optimal precision rate at 75% recall. We used Keras with TensorFlow backend to implement the convolutional neural network.

### IV. Experiments

We conducted experiments using three datasets from confidential, non-public, and real legal matters. All the documents in the three datasets were previously reviewed and received privileged or non-privileged labels from attorneys. The experiments were conducted on a list of four commonly used keyword terms in privilege review: "privi*", "legal", "counsel*", and "attorney*". Table 1. summarizes the three datasets used in our experiments. Table 2 summarizes the occurrences of the four keyword terms in the documents in the three datasets.

Table 1. Summary statistics of the three datasets

| Project | # of Documents | # of Privileged Documents | # of Non-Privileged Documents | % of Privileged Documents |
|---|---|---|---|---|
| A | 360,531 | 46,756 | 313,775 | 12.97% |
| B | 397,289 | 14,326 | 382,963 | 3.61% |
| C | 8,715,165 | 536,788 | 8,178,377 | 6.16% |

Table 2. Summary statistics of keyword term occurrences

| Keyword | Project | # of Occurrences | % of Privileged Occurrences | # of Documents Occurred | % of Privileged Documents |
|---|---|---|---|---|---|
| attorney* | A | 215,137 | 44% | 44,918 | 44% |
| counsel* | A | 257,593 | 23% | 33,842 | 38% |
| privi* | A | 256,920 | 45% | 65,557 | 41% |
| Legal | A | 603,479 | 42% | 124,951 | 29% |
| attorney* | B | 123,608 | 16% | 40,286 | 11% |
| counsel* | B | 88,254 | 15% | 30,827 | 11% |
| privi* | B | 160,349 | 17% | 74,896 | 7% |
| legal | B | 648,857 | 9% | 161,039 | 7% |
| attorney* | C | 806,926 | 51% | 268,905 | 50% |
| counsel* | C | 735,992 | 47% | 270,965 | 52% |
| privi* | C | 1,065,668 | 50% | 504,918 | 38% |
| legal | C | 3,529,572 | 51% | 907,912 | 36% |

We removed all occurrences of these four keywords in email footers from all privileged documents, because these occurrences are not privileged occurrences. In our experiments, we used the rest of all occurrences of a keyword term for each project. We trained a different CNN model for each keyword term and each project. Table 3 reports the privileged keyword occurrence precision rates at different recall levels for each dataset. The results are the averages of a five-fold cross-validation and the window size of the keyword occurrence is 20. The keyword search precisions are the precisions after keyword occurrences in email footers were removed. From Table 3, we can see the models performed well distinguishing between privileged occurrences and non-privileged occurrences.

Table 3. Precision at different recall levels for privileged keyword occurrences (%)

| Project | Keyword | Precision 75% Recall | Precision 85% Recall | Precision 90% Recall | Precision Keyword Search |
|---|---|---|---|---|---|
| A | attorney* | 95 | 91 | 85 | 33 |
| A | counsel* | 88 | 75 | 70 | 19 |
| A | legal | 96 | 93 | 89 | 42 |
| A | privi* | 97 | 91 | 87 | 31 |
| B | attorney* | 29 | 23 | 20 | 10 |
| B | counsel* | 35 | 26 | 22 | 12 |
| B | legal | 31 | 22 | 18 | 9 |
| B | privi* | 45 | 34 | 31 | 10 |
| C | attorney* | 91 | 82 | 77 | 43 |
| C | counsel* | 90 | 82 | 77 | 47 |
| C | legal | 95 | 90 | 86 | 51 |
| C | privi* | 97 | 96 | 93 | 38 |

The purpose of the study was to identify privileged documents, not the privileged keyword occurrences that exist within a document. We used the method discussed in Section 2, Learning Keyword Occurrence Models, to assign each document a score using the maximum score of all occurrences of the keyword in the document. Table 4 reports the keyword model's precisions for each project at different recall levels for all documents with at least one occurrence of the keyword.

The machine learning models provide us with the ability to trade recall for precision to achieve additional review cost savings – or to review less false positive documents and thereby save money. Assume that we want to find 75% and 90% of the privileged documents for Project A that hit on the keyword: "attorney", respectively. Only ~30% and ~40% of all documents with "attorney" hits need to be reviewed to achieve 75% recall and 90% recall. For Project B and the keyword: "privileged," only 14% of the documents need to be reviewed to achieve 90% recall. Table 5 details the document review cost savings at different recall levels.

Table 4. Precisions at different recall levels for privileged documents with an occurrence of the keyword (%)

| Project | Keyword | Precision at 75% Recall | Precision at 85% Recall | Precision at 90% Recall | Precision of Keyword Search |
|---|---|---|---|---|---|
| A | attorney* | 94 | 90 | 85 | 38 |
| A | counsel* | 94 | 87 | 79 | 36 |
| A | legal | 92 | 87 | 82 | 29 |
| A | privi* | 96 | 93 | 90 | 30 |
| B | attorney* | 36 | 25 | 22 | 8 |
| B | counsel* | 35 | 23 | 20 | 10 |
| B | legal | 25 | 16 | 13 | 7 |
| B | privi* | 47 | 29 | 24 | 4 |
| C | attorney* | 94 | 87 | 82 | 43 |
| C | counsel* | 95 | 90 | 85 | 51 |
| C | legal | 88 | 82 | 76 | 36 |
| C | privi* | 96 | 94 | 91 | 26 |

Table 5. Saving in document review at different recall levels (%)

| Project | Keyword | Saving at 75% Recall | Saving at 85% Recall | Saving at 90% Recall |
|---|---|---|---|---|
| A | attorney* | 70 | 64 | 60 |
| A | counsel* | 71 | 65 | 59 |
| A | legal | 76 | 71 | 68 |
| A | privi* | 66 | 72 | 70 |
| B | attorney* | 83 | 73 | 67 |
| B | counsel* | 79 | 64 | 55 |
| B | legal | 80 | 66 | 55 |
| B | privi* | 94 | 89 | 86 |
| C | attorney* | 65 | 58 | 52 |
| C | counsel* | 59 | 51 | 45 |
| C | legal | 70 | 63 | 58 |
| C | privi* | 80 | 76 | 74 |

We also applied a logistic regression algorithm with default parameter settings to the same datasets. Bag-of-word representation was used for logistic regression. Figure 1 Shows the precision-recall curves for the four keywords in the three datasets. In almost all cases, CNN outperforms logistic regression, but the performance improvements are not significant.

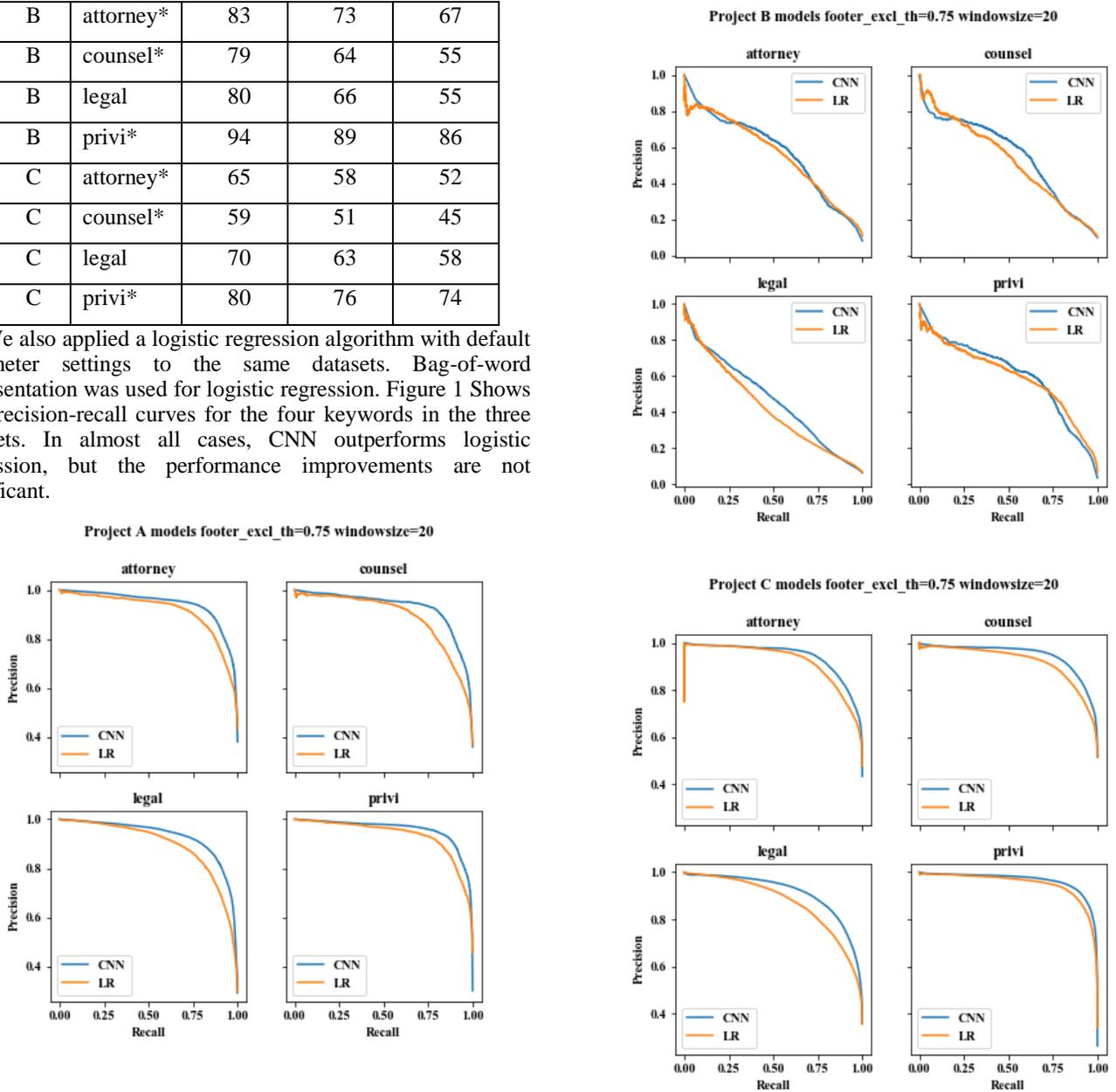

Figure 1. Performance of Comparisons of CNN with logistic regression

V. SELECTION OF TRUE PRIVILEGED KEYWORD OCCURRENCES

As we mentioned in Section II, because of the way our privileged keyword occurrences are generated, privileged keyword occurrences are only likely privileged – we know some of the privileged keyword occurrences are not actually privileged. The presence of multiple terms in a single document can create a "likely" privileged occurrence and even multiple occurrences of the same term can create uncertainty in the authenticity of a privileged occurrence. In sum, most privileged documents include multiple likely privileged keyword occurrences, some of which are false privileged occurrences, and this section outlines how we address this problem.

### A. KNN Method for Selecting Positive Examples

Recently, self-supervised learning has piqued the interest of the machine learning research community [12]. Self-supervised learning studies methods for automatically labeling training data with minimal human input and different methods have been developed for different applications [12] [13]. Most of the self-supervised learning applications were created to assist with video and image processing. The method we propose in this section is a method that can be used for labeling positive training examples. In many supervised learning applications, it is easy to find labeled negative examples, but not as easy to find labeled positive examples, for example in fraud detection applications.

Approach One – Given a set of negative examples and a set of candidate positive examples, the idea is to select candidate positive examples that are not very similar to some of the negative examples. We can use a K nearest neighbor algorithm to select a subset of candidate positive examples as positive examples. For each candidate positive example, we compute its similarity to all negative examples and select K most similar negative examples. We then assign the average similarity of the K most similar negative examples as a score to the candidate positive examples. The larger the average similarity score is, the less likely the candidate is positive. This is because true positive examples should not be similar to any of the negative examples. All candidate positive examples can be ranked using their scores in increasing order. A set of top-ranked candidates can be chosen as positive examples.

Approach Two – A more sophisticated approach involves computing the similarity between each candidate and all negative examples and also the rest of candidate positive examples. Then, K most similar negative examples and K most similar positive examples are selected, respectively and average similarities are computed for both negative examples and candidate positive examples. The score assigned to the candidate example is the ratio of the negative average similarity and the candidate positive average similarity. We can choose the set of candidates with the smallest scores as positive examples.

We implemented both methods discussed above. Our datasets are text data, so Cosine similarity was used to measure similarity. The performance was similar for both methods and we report the results of the first method (Approach One) in Section 5.2.

### B. Preliminary Experimental Results

We conducted experiments to evaluate the feasibility of Approach One using two keyword occurrence datasets from confidential, non-public, and real legal matters. The two keywords were "privileged" and "attorney" from two different projects. The dataset for "privileged" included 142,011 non-privileged occurrences and 114,909 candidate privileged occurrences from Project A. The dataset for "attorney" includes 103,264 non-privileged occurrences and 20,344 candidate privileged occurrences from Project B. The window size for the occurrences is 20, namely each keyword occurrence consists of 41 words including the keyword itself.

In our experiments, for each candidate privileged occurrence, we computed its similarities to all non-privileged occurrences and assigned the average similarity of the K (K = 3 in the experiments) most similar non-privileged occurrences to the privileged occurrence as its score. We then selected all candidate privileged occurrences with scores smaller than or equal to a cut-off score as true privileged occurrences.

In these experiments, we chosen 0.9, 0.8, 0.7, and 0.6 as cut-off scores, respectively. Table 6 shows the percentages of privileged occurrences selected for each cut-off score in both datasets. From the table, we can see similar percentages of privileged occurrences were selected for the two datasets. Many candidate privileged occurrences are very similar to some non-privileged occurrences. About 44% of the candidate privileged occurrences in the first dataset have similarities larger than 0.9, and more than 40% of the second dataset have similarities larger than 0.9.

Table 6. Percentages of privileged occurrences selected for each cut-off score in both datasets

| Similarity Cutoff Score | "Privileged" Occurrences in Privileged Documents | | "Attorney" Occurrences in Privileged Documents | |
|---|---|---|---|---|
| | Selected | Excluded | Selected | Excluded |
| 0.6 | 24% | 76% | 22% | 78% |
| 0.7 | 33% | 67% | 31% | 69% |
| 0.8 | 43% | 57% | 46% | 54% |
| 0.9 | 56% | 44% | 60% | 40% |
| 1.0 | 100% | 0% | 100% | 0% |

We randomly divided all non-privileged occurrences and selected privileged occurrences into a training set and a test set. The training set includes 70% of all occurrences, while the test set includes the rest 30%. We used SVM to build the predictive models and a bag-of-word representation with 1-gram. Stop words were not removed and no stemming was applied to the words and 2,000 words were used as features. Default parameter settings were used for SVM. Figures 2 and 3 show the ROC curves and precision recall curves for both datasets. From the figures, we can clearly see when the candidate privileged occurrences that are most similar with some non-privileged occurrences are removed, the performance significantly improved. The more candidate privileged occurrences we removed, the better the performance. Note, the distribution between the two classes becomes more unbalanced when more candidate privileged occurrences are removed.

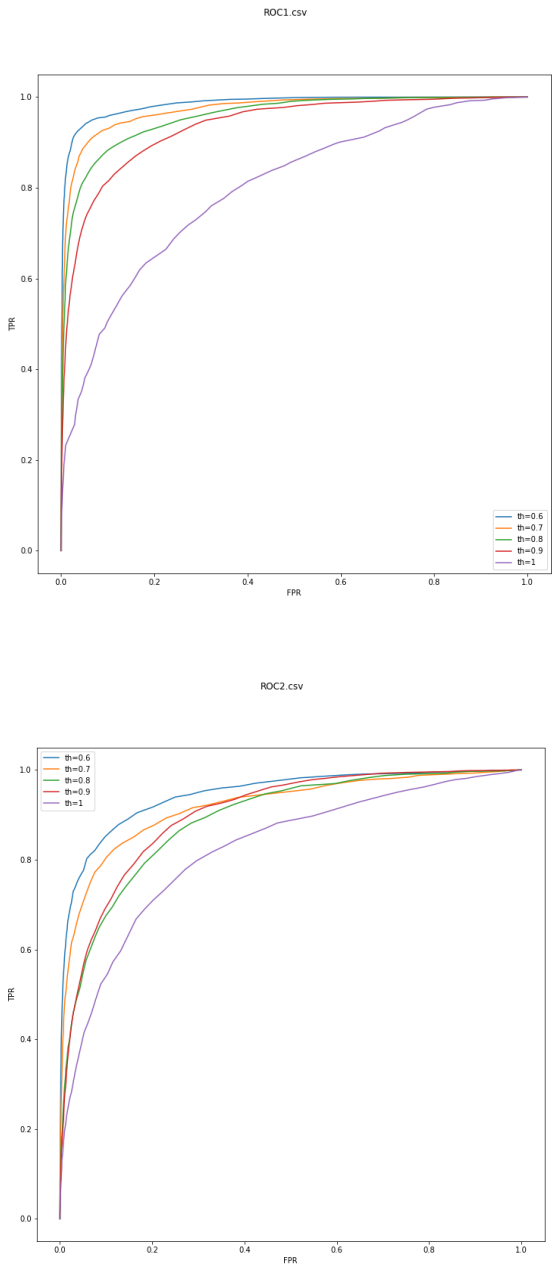

Figure 2. ROC curves for Datasets 1 and 2.

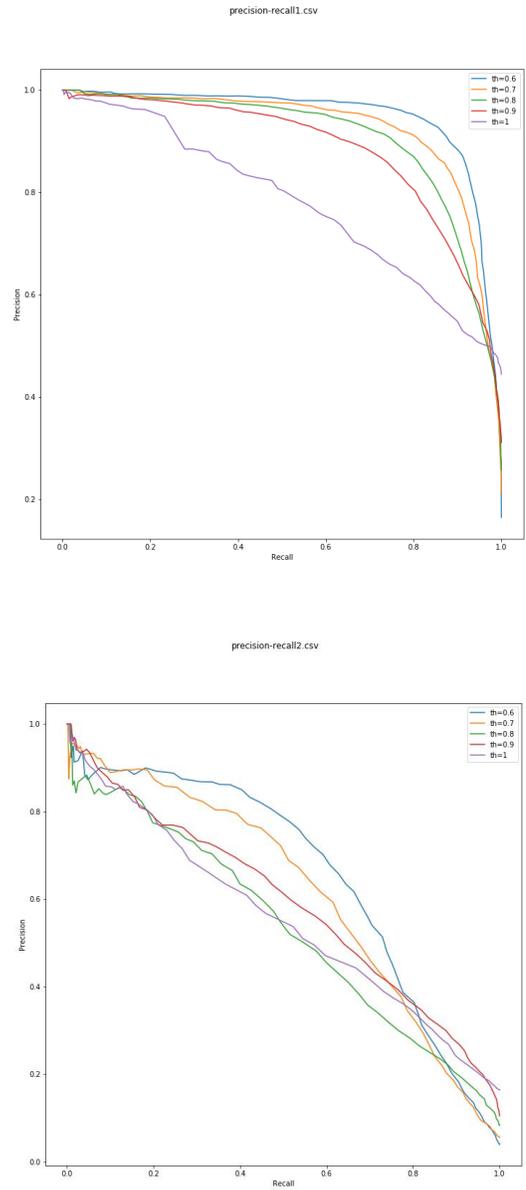

Figure 3. Precision-recall curves for Datasets 1 and 2

While promising, we should state that these performance improvements are potentially misleading. Candidate privileged occurrences were removed from our test data because they were very similar to non-privileged occurrences and some of the occurrences that we removed may be true privileged occurrences. By removing these true occurrences, we may have artificially improved the performance of the models and if we added the removed occurrences back to the test data, the recall and precision could possibly drop. In the future, we plan to evaluate the performance of the models on document level by combining all the search terms together, which can provide a more accurate evaluation of this approach's performance using the document labels.

Non-privileged occurrences in the test data remain the same for all cut-off scores. We noticed that the SVM models' scores of non-privileged occurrences became smaller when more candidate privileged occurrences were removed. Meaning, the models were getting better at detecting non-privileged occurrences. Table 7 reports the average scores of non-privileged occurrences for the SVM models learned from different sets of privileged occurrences selected with different cut-off scores.

Table 7. Average scores of non-privileged occurrences for different SVM models

|  | Dataset 1 | Dataset 2 |
| --- | --- | --- |
| Cut-off Score | Average SVM Score | Average SVM Score |
| 0.6 | 0.15 | 0.19 |
| 0.7 | 0.20 | 0.26 |
| 0.8 | 0.24 | 0.21 |
| 0.9 | 0.27 | 0.25 |
| 1.0 | 0.35 | 0.33 |

## VI. FUTURE WORK AND SUMMARY

Privileged document review is a complex requirement placed on legal teams. Determining if privileged material is present in document populations requires a nuanced calculus and, as data volumes continue to increase, remains a costly and time-consuming task. Legal teams have traditionally used search terms to target privileged materials, but search terms typically yield a large number of false positives. We proposed a machine learning approach to build models to reduce false positives generated from search terms. Preliminary results show that the proposed approach could be an effective for reducing false positives and review costs.

The ultimate goal of this research is to develop a method to build machine learning models that can be used to identify privileged documents across different legal matters. We will conduct additional studies and experiments to evaluate the feasibility in building one universal model for a keyword so that the model could be applied to identify privileged documents across different legal matters. Currently, models are matter-specific. The search term list for a legal matter could include thousands of keyword search terms and many of the search terms do not occur in a large portion of the documents. We cannot afford to build a model for each search term in this scenario. In the future, we plan to conduct research exploring the possibility of building one model for many different search terms. Lastly, we will conduct more studies in accurately selecting true privileged keyword occurrences.